\documentclass[aps,prb,twocolumn,superscriptaddress,showpacs]{revtex4}
\usepackage{graphicx}
\usepackage{epsfig}
\usepackage{bm}
\usepackage{times}
\usepackage{amssymb}
\usepackage{url,hyperref}

\begin{document}

\title{Magnetic excitations in vanadium spinels}

\author{N. B. Perkins}
\affiliation{Institute fur Theoretische Physik, TU Braunschweig,
Mendelssohnstrasse 3, 38106 Braunschweig, Germany}
\affiliation{Bogoliubov Laboratory of Theoretical Physics, JINR,
Dubna, Russia}
 \author{O.  Sikora}
\affiliation{MPIPKS, N\"othnitzer Str. 38, 01187 Dresden, Germany}

\date{\today}

\begin{abstract}

We study magnetic excitations in vanadium spinel oxides AV$_2$O$_4$
(A=Zn, Mg, Cd) using two models: first  one is a
 superexchange model for vanadium $S=1$ spins, second one includes in
addition spin-orbit coupling, and crystal anisotropy. We show that
 the experimentally observed magnetic ordering can be obtained in both
models, however the orbital ordering is different with and without
spin-orbit coupling and crystal anisotropy. We demonstrate that this
difference strongly affects the spin-wave excitation spectrum above
the magnetically ordered state, and  argue that the neutron
measurement of such dispersion
 is a way to distinguish between the two possible orbital orderings in
AV$_2$O$_4$.

\end{abstract}


\maketitle

\section{Introduction}

Due to geometrical frustration, transition metal spinel oxides
 with a general formula $AB_2$O$_4$  display  a variety of unusual
low-temperature properties. The spin dynamics of these systems is
usually described by a Heisenberg antiferromagnet  on the
pyrochlore lattice. This model  is rather peculiar and its
 classical ground state is  highly degenerate. Exact degeneracy can be lifted
 by various mechanisms, but the system still possesses
 many competing spin configurations
 with almost equal  energies. As a result, when the temperature goes down,
the system can evolve in a variety of ways: it can remain  spin
liquid down to the lowest temperatures due to quantum fluctuations,
or choose a particular configuration either via the strongest
 order from disorder mechanism or through a
structural phase transition which lowers the local symmetry of the
lattice.

A peculiarity of transition metal spinels is that their  magnetic
ions often possess also an orbital degree of freedom. This extra
degree of freedom modulates the spin exchange and can at least
partially lift the geometrical degeneracy of the underlying lattice.
 However, the orbital degrees of freedom in geometrically frustrated
lattices are by themselves  frustrated, i.e.,
 many different orbital configurations have the same energy.
In this situation, the ordering in the orbital sector is coupled to
the
 ordering in the spin sector, and the selection of a true ground state
 configuration becomes a non-trivial  phenomena.

Of  particular interest is the orbital ordering  in the $t_{2g}$
systems on the pyrochlore lattice. In this work we study vanadium
spinels of the type $A$V$_2$O$_4$, where $A$ is a divalent ion like
Cd$^{2+}$, Zn$^{2+}$, or Mg$^{2+}$. In these compounds, magnetically
active V$^{3+}$ ions form a pyrochlore lattice and have two $3d$
electrons in $t_{2g}$-orbitals. Due to strong Hund's interaction,
these two electrons form a state with $S=1$.  All $A$V$_2$O$_4$
compounds show qualitatively similar structural and magnetic
behavior, independently of what the divalent A ion is, and  undergo
two phase
 transitions -- a structural one and an antiferromagnetic  one.
We will be mainly discussing the physics of ZnV$_2$O$_4$. The structural
transition occurs at a temperature $T_S \sim $~50~K.~\cite{10tcher}
Below $T_S$,  the lattice
shows a tetragonal distortion --  the vanadium octahedra VO$_6$ are
uniformly flattened along the $c$ axis, and
 the symmetry  is lowered   from the cubic one to
 $I4_1/amd$, which is the highest-symmetry tetragonal space group for the
 spinel  structure.
  The  antiferromagnetic (AFM)
transition occurs at a slightly lower temperature $T_N$ of about
40~K.~\cite{niziol,reehius} This temperature is significantly lower
than the Curie-Weiss temperature $T_{CW} \sim 1000K$, extracted from high
temperature susceptibility,~\cite{Muhtar} which underlines
 the importance of geometrical frustration.

The magnetic structure of AV$_2$O$_4$ spinels at $T < T_N$ was
 first proposed by Nizio\l~\cite{niziol} and
recently confirmed by Reehius {\it et al.} in Ref.~\onlinecite{reehius}. Along the
diagonal $[110]/[{\bar 1}10]$ direction in the $xy$-plane, the
ordering is antiferromagnetic + - + -..., while along the two other
diagonal directions $[011]/[0{\bar 1}1]$ in  the $yz$-plane and
$[101]/[{\bar 1}01]$ in the $xz$-plane, the spin ordering  is in the
form + + - - + + - -... (see Fig.~\ref{fig1}).

At high temperatures $T > T_S > T_N$,
inelastic neutron scattering data by Lee {\it et al.}~\cite{lee} on
the powder sample of ZnV$_2$O$_4$ still show  strong low energy magnetic
excitations which form  a broad peak centered at $Q=1.35~$\AA$^{-1}$.
This broad peak is  present also at  $T_N<T<T_S$, however it  becomes
asymmetric and shifts towards a smaller value of $Q$.
 The asymmetry of the
peak further increases in the magnetically ordered phase
($T<T_N$).~\cite{lee} The value of $Q$ and its temperature evolution
cannot be explained within a purely spin model, whose  fundamental
degrees of freedom
 are antiferromagnetic hexagonal  spin loops with $Q=1.5~$\AA$^{-1}$,
 which is larger than the experimental $Q$. The spin dynamics, however, can
be understood if one assumes that spin degrees of freedom are
affected by the orbital degrees of freedom.
 In particular, the spatial asymmetry of the peak in the  intermediate
phase  can be understood as the consequence of the fact that at
 $T < T_S$, vanadium octahedra are flattened and
$xy$-orbital at each site is occupied. This leads to a strong
antiferromagnetic exchange between vanadium spins along $xy$
direction, and, as a result, spin interactions become effectively
one-dimensional. Strong one-dimensional spin fluctuations give rise, via
spin-orbit coupling, to fluctuations of the  occupations of
$xz$ or $yz$ orbitals, causing the anisotropy of the neutron peak.

There were several theoretical attempts to understand the nature of
the ground state of ZnV$_2$O$_4$. However, although it is widely
accepted that  orbital degrees of freedom play an important role, no
consensus is reached yet about the type of the orbital ordering (OO)
in the ground state.
 The first attempt to explain the physics of
ZnV$_2$O$_4$ focused on the spin-lattice coupling
mechanism,~\cite{yam} however it didn't explain
 why the structural and spin order occur at different
temperatures. Tsunetsugu and Motome~\cite{tsun} later addressed this
 issue and
related the presence
of two separate phase transitions at $T_N$ and $T_S$ to the
interplay between geometrical frustration and
 $dd\sigma$ superexchange  (SE) interaction
between V-ions.  The ground state orbital ordering suggested in
Ref.~\onlinecite{tsun} consists of stacked $ ab$ planes with
alternating orbital occupations $(xy,xz)$ and
$(xy,yz)$. Hereafter  we label this orbital patterns
as ROO. They also showed that this ordering of orbitals can partially
remove magnetic frustration and explain experimentally observed
 ordered magnetic structure.

On the other hand, Tchernyshyov~\cite{tcher} pointed out that the
ground state obtained in Ref.~\onlinecite{tsun} is at odds with
x-ray and neutron diffraction data, because it does not possess the
required $I4_1/amd$ space symmetry. He argued that the spin-orbit
coupling should be included into consideration. He considered a
purely ionic model in which spin-orbit (SO) coupling plays the major
role and determines the orbital order in the tetragonal phase.  He
suggested  the following OO: one electron on each site occupies
$xy$ orbital, while the second electron is spreaded between
 $xz$ and $yz$ orbitals in such a way as to minimize the spin-orbit energy at each site. Hereafter we label this
orbital pattern as COO.

Recently R. Valenti {\it et al.}\cite{valenti} found in the
ab-initio DFT calculations that a correct space symmetry can be
actually obtained within the reasoning of Ref.~\onlinecite{tsun}, if one
includes into
 consideration an additional trigonal distortion.
Still, the true ground state turns out to be the same as in
Ref.~\onlinecite{tcher}.

The ideas of Refs.~\onlinecite{tsun} and \onlinecite{tcher} were
combined in the unique framework by S. Di Matteo {\it et
al.} in Ref.~\onlinecite{perkins1}. They proposed to construct
a classical ground state phase diagram by considering
SE interaction and SO coupling  on equal footing. They demonstrated that
 the SO  coupling is  a relevant perturbation and
favors the states with unquenched orbital momentum for any value of
the coupling strength.  They obtained a variety of phases and found
that for reasonable values of SE and SO couplings, the ground state agrees
with the experimentally observed one.

In this paper we extend the analysis of Ref.~\onlinecite{perkins1}
and study the  low energy excitations in vanadium spinels.   We find
that the magnetic excitation spectrum strongly depends on the type
of the OO, and  that it is qualitatively different for the ground
states  with  patterns that consist of  real orbitals  and those
with the  complex linear combination of orbitals,  i.e. complex
orbitals. The former ground state is characterized by a quenched
orbital angular momentum ($ L =0$), while the later  by a unquenched
($L\neq 0$) orbital angular momentum.~\cite{extra} The difference in
the magnetic excitation spectrum arises due to the fact that the
magnetic moment of the vanadium ion is formed by both spin and
orbital momentum, and fluctuations of ${\bf L}$
 contribute to the spectrum of magnetic excitations.
 We argue that the  measurement of  magnetic excitations
in neutron scattering experiments  can shed light on the
nature of the OO in the ground state.

This paper is organized as follows: in Section II we introduce
 the  model appropriate for the  description of the
 physical properties of ZnV$_2$O$_4$.
 In Section III we discuss  the ground state  and the magnetic excitations
 of the system when the orbital angular momentum is quenched. We derive
 linear spin-wave theory of $S=1$ moments interacting on the $V^{3+}$
pyrochlore lattice. In Section IV we discuss ground state and
magnetic excitations of the system when the orbital angular momentum
is unquenched. We show that magnetic excitations for unquenched
orbital angular momentum
 can be described in the framework of the
magnetic excitonic model.  Section V presents the conclusions.
 Some mathematical details are given in the Appendices A and B.

\section{The Model}

The minimal  model  describing the low energy physics of
 vanadium spinel   is given by
\begin{eqnarray}
 \begin{array}{l}
H=H_{\rm SE}+H_{\rm a}+H_{\rm SO}~.
\label{hamil}
\end{array}
\end{eqnarray}
\noindent The  first term describes  nearest neighbors ($nn$)
super-exchange interactions between vanadium $S=1$ spins,
arising from the virtual excitations $d_i^2d_j^2\rightarrow
d_i^1d_j^3$.  These interactions can be
written as:
\begin{eqnarray}
\begin{array}{ll}
H_{\rm SE}^{nn}= &-\sum_{<ij>}{\big [}J_0 {\bf S}_i\cdot {\bf S}_{j}+
J_1{\big ]} O_{ij} \\
 &-\sum_{<ij>}J_2{\big [}1-{\bf S}_i\cdot {\bf S}_{j}{\big ]}{\bar
O}_{ij}~,
\end{array}
\label{spinorb2}
\end{eqnarray}
where $i$ and $j$ are nearest neighbors,
 $J_0=\eta J/[1-3\eta]$, $J_1=J[1-\eta]/[1-3\eta]$,
$J_2=J[1+\eta]/[1+2\eta]$ are coupling constants, $J=t^2/U_1$  is
the overall energy scale ($t=3/4\,t_{dd\sigma}$
 and $U_1$ is the intra-orbital Coulomb repulsion),
  $\eta =J_H/U_1$ is the normalized Hund's exchange.
 We consider only the largest the hopping term, associated with
 $\sigma$-bonding~\cite{ddsigmaproof}. Such hopping is
diagonal and non-zero only if the orbitals and the plane in which
hopping occurs  are of the same $\alpha\beta$ type ($\alpha\beta=xz$,
$yz$, $xy$).
In this case orbital contributions $O_{ij}$ and ${\bar O}_{ij}$ are
expressed in terms of projectors $P_{i,\alpha\beta}$ onto the
occupied orbital state $\alpha\beta$ at site $i$ and $j$: $O_{ij}=
P_{i,\alpha\beta}(1-P_{j,\alpha\beta})+
P_{j,\alpha\beta}(1-P_{i,\alpha\beta})$ and
$\bar{O}_{ij}=P_{i,\alpha\beta}P_{j,\alpha\beta}$.

To describe the anisotropy and  spin-orbit coupling  term, we
use the fact that, when the crystal field splitting between
$t_{2g}$ and $e_g$ orbitals is large, the $t_{2g}$-electrons can be
represented by an effective orbital angular momentum
$L'=1.$~\cite{ball} The anisotropy term (the second
term of the Hamiltonian  (\ref{hamil})),  is then given by
\begin{eqnarray}
H_{\rm a}=c\sum_{i}L^{'2}_{zi}~.
\label{anis}
\end{eqnarray}
\noindent where $c$ is a constant. This term  describes the
tetragonal distortion in the $t_{2g}$ manifold. We notice, that here
for simplicity we neglect the trigonal distortion, which is small
compared to the tetragonal one.

The  spin-orbit coupling term ( the third term in (\ref{hamil}))
 is given by
 \begin{eqnarray}
H_{\rm SO}=-\lambda\sum_{i}{\bf L}'_i\cdot\bf{S}_i~, \label{so}
\end{eqnarray}
\noindent where $\lambda$ is SO coupling constant. Note that the
true angular momentum $\bf{L}$  is related to an effective one as
$\bf{L}\simeq-\bf{L}'$.

The parameters of the model can be estimated from the experiments.
The spectroscopy data~\cite{V} yield the
 Hund's exchange, $J_H \simeq 0.68$ eV and Coulomb intra-orbital
 repulsion $U_1\simeq 6$~eV. The estimate of hopping matrix element from
x-ray photoemission spectroscopy is $t\simeq
-0.35$~eV.\cite{ddsigmaproof}, so the  energy scale is
$J=t^2/U_1\simeq 20.4$~meV. Since the SO coupling constant is
$\lambda\simeq 13$~meV (Ref.~\onlinecite{abrag}),
 we can see
that the super-exchange and  the spin-orbit couplings are comparable
and, therefore should be treated on equal footing.

\section{Real orbital order}

\subsection{Ground state}
First we discuss the ground state  of the super-exchange Hamiltonian
alone (Eq. \ref{spinorb2}).  Without the anisotropy and spin-orbit
terms the ground state orbital patterns consist of only real
orbitals: at each site two out of three $t_{2g}$ orbitals
$xy,~xz,~yz$, are occupied. This type of orbital patterns  is called
real orbital order, as opposed to complex orbital order, when the
orbital state is formed by a complex superposition of $t_{2g}$
orbitals in such a way, that the gain of spin orbit interaction
energy is maximized.

Depending on which orbitals are occupied, one obtains two types of
interacting bonds and also non-interacting bonds. Consider for example a
bond in the $\alpha\beta$-plane. If there is an
electron occupying $\alpha\beta$ orbital only on
 one site of such bond, then the bond, which we label as $b_1$,
is weakly ferromagnetic (FM), and is described by
$H_{b_1}=-J_{0}{\bf S}_i\cdot {\bf S}_{j}-J_1$. If both $i$ and $j$
sites of $ij$ bond are occupied by $\alpha\beta$ electrons,
then the bond, which we label as $b_2$, is strongly antiferromagnetic (AFM).
 The exchange coupling
is then given by $H_{b_2}=-J_2(1-{\bf S}_i\cdot {\bf S}_{j})$. When
neither $i$ nor $j$ site have $\alpha\beta$-orbital occupied, the
bond is non-interacting.

 One can easily demonstrate~\cite{perkins1}
that for positive $J_0$ and $J_2$, the  lowest energy configuration
corresponds to the state with four ferromagnetic $b_1$ bonds per
tetrahedron. There still exist two topologically different
tetrahedral configurations
 with  four ${b_1}$ bonds,
characterized by different OO patterns. One of them, the ROO state
with OO patterns proposed by Tsunetsugu ~\cite{tsun}, is compatible
with experimentally observed magnetic structure -- it yields AFM
chains running in [110] and [1$\bar{1}$0] directions (see,
Fig.\ref{fig1}). The classical energy per site in the ROO  state is
$E_{\rm ROO}=-2J_1-2J_2$. However, this state is actually not the
true ground state of the super-exchange Hamiltonian
(\ref{spinorb2}). The other state with 4~${b_1}$ bonds (the ROO-I
state in our notations),
  in which
spins of each tetrahedron form a fully collinear up-up-up-down
($uuud$) state, has the ground state energy
 $E_{\rm ROO-I}=-J_0-2J_1-2J_2$, which is
 lower than that of the ROO state.  This is the lowest classical
energy that one can obtain within the manifold with quenched angular
momentum at each site~\cite{perkins1}. However, the magnetic
ordering associated with the ROO-I  state is incompatible with the
experimentally observed one: in ROO-I two neighboring spins in
$xy$-plane are ferromagnetically aligned whereas the experimentally
 detected coupling in $xy$-plane is antiferromagnetic.
This discrepancy  between the ROO-I ground state  of the super-exchange
Hamiltonian ({\ref{spinorb2})  and the experimental findings
demonstrates the necessity of taking into account additional interactions
We study the the anisotropy term and the spin-orbit coupling in the Section IV. \\
One can also consider the model with both the super-exchange (SE)
and the Jahn-Teller  (JT) couplings \cite{motome} with a hope that
the cooperative JT effect, which plays an important role in the
structural transition, can also stabilize the ROO type of orbital
order. This model has been studied in Ref.~\onlinecite{motome} by
both the mean-field analysis and Monte Carlo simulations. Here we
assume that the ROO phase can be realized and in the next subsection
we derive magnetic excitations spectrum for the corresponding
effective spin model.

\begin{figure}
\vspace{-2.5cm} \rotatebox{270}{
\centerline{\epsfig{file=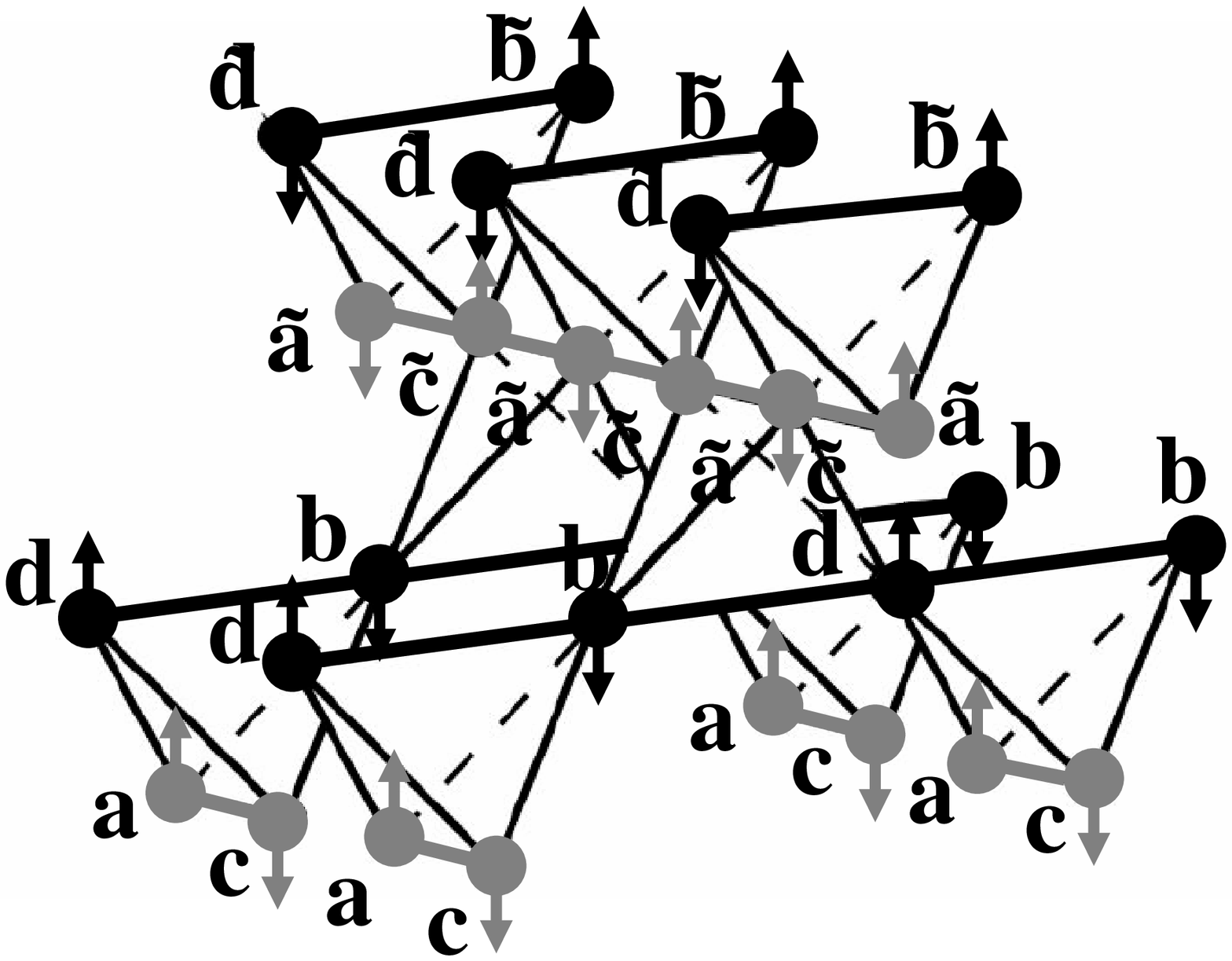,width=4cm}}} \vspace{-2cm}
 \caption{(Color online) Magnetic
ordering consisting of one-dimensional antiferromagnetic chains in
the $xy$-plane. Black and grey  colors correspond to the orbital
configuration in the ROO state.} \label{fig1}
\end{figure}

\subsection{Spin waves}

The starting point for the calculation of the magnon excitation
spectrum is the classical Neel ground state with antiferromagnetic
spin chains in $xy$-planes (Fig.~\ref{fig1}).  This ground state has
the magnetic unit cell with eight  vanadium spins, which we denote
as
 $a,b,c,d,\tilde a, \tilde b,\tilde c,\tilde d$ (Fig.\ref{fig1}).
After averaging the orbital operators we can re-write the Hamiltonian
(\ref{spinorb2})  as
\begin{equation}
H_{nn} = J_{xy}\sum_{\langle ij \rangle\parallel xy}
{\bf S}_i\cdot {\bf S}_j+
J'\sum_{\langle ij \rangle\parallel xz,yz}
{\bf S}_i\cdot {\bf S}_j
\label{hamunit}
\end{equation}

\noindent where the first term describes the super-exchange along
$xy$-chain with $J_{xy}=J_{2}$,
while the second term corresponds to the
frustrated ferromagnetic inter-chain coupling, $J'=J_0$.\\
To describe the excitation spectrum of such eight-sublattice
antiferromagnet,  we introduce eight  boson operators:
$a,b,c,d,{\tilde a},{\tilde b},{\tilde c},{\tilde d}$. We employ
Holstein-Primakoff transformation, e.g. for up-spins ${{\bf S}}_a$
 and down-spins ${{\bf S}}_b$ we have:
\begin{eqnarray}
\begin{array}{ll}
\label{eaz}
S_a^z = S- a^{\dag}a &~~~ S_b^z = -S+ b^{\dag}b          \\
S_a^+ = \sqrt{2S-a^{\dag}a}~a &~~~ S_b^+ = b^{\dag}\sqrt{2S-b^{\dag}b}\\
S_a^- = a^{\dag}\sqrt{2S-a^{\dag}a}&~~~S_b^- = \sqrt{2S-b^{\dag}b}~b\\
\end{array}
\end{eqnarray}
\noindent
In the linear spin wave approximation we
 substitute $ \sqrt{2S-p^{\dag}p}= \sqrt{2S}$ in the expressions above.
 Performing Fourier transformation
 $p_k=\frac{1}{\sqrt{N}}\sum_i \exp^{-\imath\vec{k}\vec{r_i}}p_i$,
 where $N$ is the number of lattice sites belonging to one sublattice,
we obtain the mean field Hamiltonian for magnons~\cite{walker}
\begin{equation}
H =({\bf a}^{\dagger}(\bf k), -{\bf a}(-\bf k))
\left(\begin{array}{cc}
{\bf A}({\bf k})  &  {\bf B}({\bf k})  \\
  -{\bf \tilde B}(-{\bf k})  &  -{\bf \tilde A}(-{\bf k})
\end{array} \right)
\left(\begin{array}{c}
 {\bf a}({\bf k}) \\ {\bf a}^{\dagger}(-{\bf k})
\end{array} \right),
\label{eakoma}
\end{equation}
\noindent where we introduced ${\bf a (\bf
k)}=(a_{\bf k},b_{\bf k},c_{\bf k},d_{\bf k},{\tilde a}_{\bf
k},{\tilde b}_{\bf k},{\tilde c}_{\bf k},{\tilde d}_{\bf k}$). The matrices
 ${\bf A}({\bf k})$ and ${\bf B}({\bf k})$ are eight by eight matrices whose
elements depend on the geometry of the lattice and the type of
the magnetic ordering, tilde denotes  the complex conjugation. The
explicit expression for matrix elements are presented in Appendix A.
The quadratic form is diagonalized using the generalized Bogoliubov
transformation. In the diagonal form, the
Hamiltonian takes the form:
\begin{eqnarray}\nonumber
H_{{\bf k}, -{\bf k}} = H_{{\bf k},-{\bf k}}^0+ \sum_{n}
\lambda_{n\bf k}\,\, b_{n\bf k}^{\dagger}\, b_{n\bf k}\\ +\sum_{n}
\lambda_{n\bf -k}\,\, b_{n\bf -k}^{\dagger}\, b_{n\bf -k}
\label{hdiag}
\end{eqnarray}
where the index $n$ runs from 1 to 8,  $\lambda_{n\bf k} =
\lambda_{n\bf -k}$ are magnon energies and $b_{n\bf k}$ are linear
combinations of boson operators belonging to $\bf a(\bf k)$ and $\bf
a(-\bf k)$.

We obtained the spin-wave excitation spectrum by solving
Eq. (\ref{hdiag}) numerically. The result
 is presented in Fig.~\ref{fig2}}  along
high-symmetry directions of 3D Brillouin zone. We used $J_{xy}=18.5$
meV, and the ferromagnetic constant $J' = -0.1\,J_{xy}$. There are 4
different branches of the spin-wave spectrum, each of them is doubly degenerate.
Furthermore, two of the branches
 have zero energy over a finite range of momenta  (see
Fig.~\ref{fig2} (solid lines)). This  so called zero modes
 emerge because the number of ferro- and
antiferro-bonds connecting two neighboring antiferromagnetic chains
is the same (see Fig.~\ref{fig1}), and  the spins forming $xy-$chains can
collectively rotate with no change in energy.

The existence of the zero modes is inconsistent with the
 observation of the magnetic ordering transition at $\sim 40$K.
The ordering requires that the zero mode be lifted. The issue is
what interactions are responsible for the lifting of spin
degeneracy.

\begin{figure}
\centerline{\epsfig{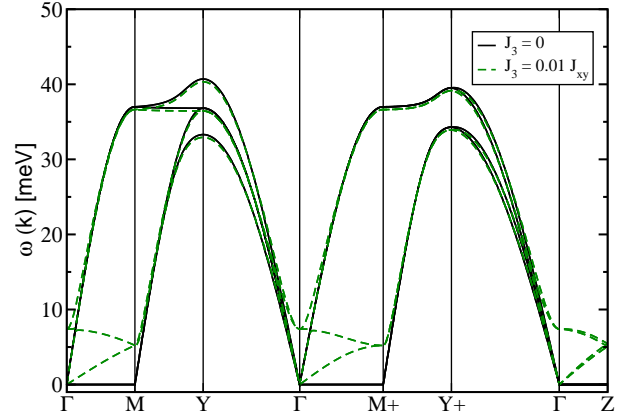}}
 \caption{(Color online) The spin-wave dispersions obtained along the main directions
 of the 3D BZ. Solid line corresponds
 to the magnon spectrum when  $J_3=0$, dashed line -
 to the magnon spectrum with the correction due to the  third nearest neighbor
 interaction with $J_3=0.01\,J_{xy}$ .
 We have used the following  labels for  high-symmetry points:
$\Gamma =(0,0,0)$, $M=(\pi/4,\pi/4,0)$,
 $Y=(0,\pi/2,0)$, $M+=(\pi/4,\pi/4,\pi/4)$,
 $Y+=(0,\pi/2,\pi/4)$, $Z=(0,0,\pi/4)$. } \label{fig2}
\end{figure}

A natural first step would be to consider  longer range
interactions~\cite{motome} as these interactions generally remove
the degeneracy (it happens, e.g., in a Kagome antiferromagnet).
 We show below that it is indeed the case, however, the
 energy of the relevant degeneracy-breaking mode is very small and
  can not explain the magnetic ordering temperature of $\sim 40$K.

\begin{figure}
\centerline{\epsfig{file=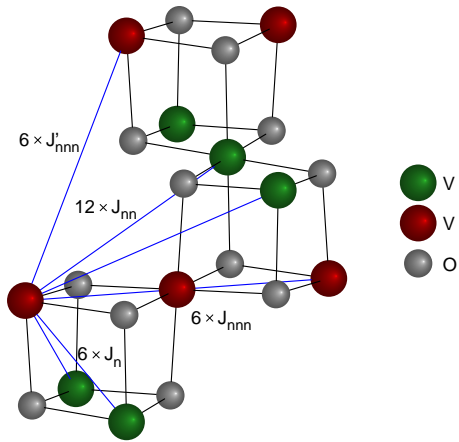,width=6cm}}
 \caption{ (Color online) $J_{n}$, $J_{nn}$, and $J_{nnn}$ are first-, second- and
 third-neighbor exchange couplings. Second-neighbor
interactions $J_{nn}$ are frustrated.
 The third-neighbor exchange coupling through the empty space  we assume
 equal to  zero,
 $J'_{nnn}=0$; the other coupling are equal to $J_{nnn}=J_3$, only if
  they connect sites along the direction corresponding to the symmetry of
  orbital occupation.
 Red and green colors denoting vanadium  ions correspond to
orbital
 configurations in ROO state.
 }
  \label{fig2a}
\end{figure}

In Fig.\ref{fig2a} we show the interactions which include up to
third neighbors. There are 6 nearest neighbor interactions $J_n$
($J_n=J_{xy}$ along $xy$ chain, and $J_n=J'$ along $xz$ and $yz$
bonds), 12 second neighbor interactions $J_{nn}$, and 12 third
neighbor interactions $J_{nnn}$. Quite often already inclusion of
the second neighbor exchange  lifts the degeneracy. However, here
second-neighbor interactions $J_{nn}$ are frustrated and can not
remove the degeneracy, and, therefore,  zero modes in the spin-wave
spectrum.~\cite{motome} Thus, one  has to include third neighbor exchanges.
There are two inequivalent sets of
 third neighbors, one  obtained by two nearest neighbor steps $J_{nnn}$ and
the other through the empty space $J'_{nnn}$ (see Fig.\ref{fig2a}).

 When only $dd\sigma$ hopping  is considered,
 the exchange coupling through the empty space is zero, $J'_{nnn}=0$,
 and only $J_{nnn}$ interactions contribute.
  These interactions are antiferromagnetic  $J_{nnn}=J_3 >0$,
 and non-zero only if  they connect sites along
 the direction corresponding to the symmetry of  the
orbital occupation, (i.e., for orbital occupation $\alpha\beta$, the
interaction is nonzero only along $\alpha\beta $ direction).

The third neighbor interaction is frustrated along $xy$-chains, but
it is  small compared to nearest neighbor  exchange along the chain,
$J_3\ll J_{xy}$, and can not destroy antiferromagnetic ordering
along the chain. Along $xz$ and $yz$ directions, $J_3$ are not
frustrated, and connect parallel antiferromagnetic chains located in
second neighboring $xy$ planes.
 The energy
scale for $J_{3}$ is then $\frac{t_{dd\sigma}^4}{U_1^3}$. This is a
very small energy,
 only about one percent of the frustrated ferromagnetic inter-chain coupling
$J'\sim \frac{t_{dd\sigma}^2}{U_1}$. This small
interaction can not explain the magnetic ordering temperature of
$\sim 40$~K.~\cite{notes1}  Here we consider this interaction only
qualitatively and assume the value of coupling constant
$J_3=0.01\,J_{xy}$. One can see in Fig.~\ref{fig2} (dashed lines) that
the zero energy modes indeed become dispersive for $J_3 > 0$.

We would like to note that experimentally
it has been proven that all magnetic moments are aligned along the
$z$-direction. This experimental fact cannot be explained in the
framework of the SE model, because it is isotropic in a spin space.
In reality, vanadium spinels
likely possess a single ion spin anisotropy which aligns spin along
the $z$-direction. The anisotropy does  affect the magnetic
excitation spectrum which in its presence  acquires  a gap.
In this case, the zero mode will be lifted
and the spin frustration will be removed. However, since the
strength of magnetic anisotropy is not known
 experimentally at the moment, we cannot estimate the magnitude of the
anisotropy-induced  gap and check whether or not this interaction
 alone can stabilize the  ground state magnetic structure.
We believe that the role of the anisotropy deserves further
 experimental and theoretical investigation.

\section{Complex orbital order}

\subsection{Ground state}

We now consider the ground state of the system in the presence of
the SO coupling and the anisotropy term (Eqs. (\ref{anis}) and
(\ref{so})). For any finite $\lambda$ the SO coupling prefers the
orbital state with the unquenched effective orbital angular momentum
$L'=1$. In such orbital state, one electron at each site
 occupies $xy$ orbital due to the
tetragonal distortion, while the second electron occupies the
complex linear combination of $xz$ and $yz$ orbitals.
 The
effective ${\bf L}^\prime$ should then be parallel to the spin
magnetic moment  in order
 to minimize  the spin-orbit energy, i.e.  a spin-up site will have
 $L'_z=1$ while a spin-down site will have $L'_z=-1$.
 As we discussed earlier, this type of orbital ordering is a COO state suggested
first by Tchernyshyov in Ref.~\onlinecite{tcher}. Its energy
$E_{COO}=-1/2[5J_2+2J_1]-\lambda$ is lower than the energy of the
ROO state  for a wide range of parameters (see the phase diagram in
Ref.~\onlinecite{perkins1}).

The COO state is characterized by two strong AFM bonds  per
tetrahedra, and its magnetic structure consists of AFM chains in
$xy$-planes with the same  interaction along the chain as in the ROO
state. The strengths of the inter-chain coupling $J^\prime$  are
also practically the same for the ROO and the COO
states~\cite{perkins1}, although
 $J'=1/4[J_2-2J_0]$ is antiferromagnetic in the COO state,
while it is ferromagnetic in the ROO state. In this work we assume
$|J'|/J_{xy} = 0.1$ in both orbital states. In the COO state the preferred
spin direction is fixed by  anisotropy term (\ref{anis}) via the SO coupling
to be along the $z$-axis, and, therefore, spins in $xy$-chains cannot
rotate freely even for only nearest-neighbor exchange along the
chains.  The long range order + + - - ... along diagonal directions in
$xz$ and $yz$ planes cannot be determined by local interactions, but
for simplicity we do not include next neighbor hopping terms in this
part of our calculations.

\subsection{Magnetic excitons}

We now consider  magnetic excitations in the COO state.
 We  follow the  magnetic exciton model approach  of
Refs.~\onlinecite{buyers,tomiyasu}, which is the extension of the
linear spin wave theory for
 systems with unquenched orbital angular
  moment.

We consider states with the effective total angular momentum ${\bf J}
= {\bf L}'+{\bf S}$. Often, energy levels with different $J$ are
well separated in energy and both  $J$
 and its $z$-projection $J_z$ are good quantum numbers.
 However, in many transition metal oxides
the strength  of the spin-orbit coupling and the super-exchange
interaction between localized $d$-electrons are comparable, and
atomic energy levels with different values of $J$
 can cross each other; in
such case only  $J_z$ acts as a good quantum number. We show below
that  magnetic excitations in vanadium spinels AV$_2$O$_4$ can be
understood  as a propagation of excitations to states with a given
$J_z$ through the crystal.

To proceed, we rewrite the Hamiltonian (\ref{hamil}) as a sum of a
single-ion Hamiltonian $H_1$ and
 the term which describes the interaction between two different ions $H_2$:
 \begin{eqnarray}
H=H_1+H_2~, \label{h12}
\end{eqnarray}
\noindent where
 \begin{eqnarray}
 \begin{array}{l}
H_1=H_{\rm SO}+H_{\rm a}+ \sum_{i}h_{zi}S_{zi}~, \label{h1}\\
H_2=H_{\rm SE}-\sum_{i}h_{zi}S_{zi}~.
\end{array}
\label{h2}
\end{eqnarray}
 \noindent
The molecular field part of the exchange interaction acting on site
$i$ is given by $h_z=\sum_{r}Z_rJ_r\langle S_z\rangle_r$. $Z_r$ is
the number of $r$-th neighbors, $J_r$ is the corresponding exchange
constant, and $\langle S_z \rangle$ is the sublattice magnetization.

First, we diagonalize the single ion Hamiltonian, $H_1$, in the
molecular field approximation. It is convenient to express the
eigenfunction $|J_z\rangle$  for the states splitted by the
spin-orbit interaction
 as linear combinations of the unperturbed eigenfunctions of
$L_z'$ and $S_z$. Then
 $H_1$ can be represented as a block $9\times9$ matrix
in the subspace of $|L_z',S_z\rangle$ as follows:
\begin{eqnarray}
H_1(J_z=\pm 2)=c-\lambda \pm h_{z} \label{jz2}
\end{eqnarray}
\begin{eqnarray}
H_1(J_z=\pm 1)= \left(\begin{array}{cc} c&-\lambda\\
 -\lambda&
\pm h_{z} \end{array}\right)
 \label{jz1}
\end{eqnarray}
\begin{eqnarray}
H_1(J_z=0)= \left(\begin{array}{ccc} c+\lambda
-h_z&-\lambda&0\\
 -\lambda&0&-\lambda\\
0&-\lambda&c+\lambda +h_{z} \end{array}\right)
 \label{jz0}
\end{eqnarray}
\noindent where  $J_z=2$, $J_z=-2$, $J_z=1$, $J_z=-1$ and $J_z=0$
are represented in the basis $|1,1\rangle$, $|-1,-1\rangle$,
($|1,0\rangle$, $|0,1\rangle$), ($|-1,0\rangle$, $|0,-1\rangle$),
and ($|1,-1\rangle$, $|0,0\rangle$, $|-1,1\rangle$), respectively.
Diagonalizing these matrices, we obtain eigenvalues and eigenvectors
of  $H_1$ as  functions  of the molecular field $h_z$.

\begin{figure}
\centerline{\epsfig{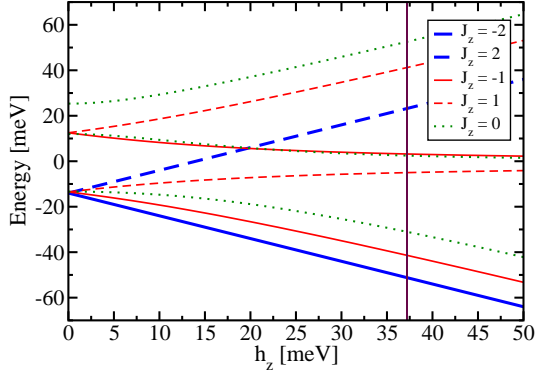}} \caption{ (Color online)
Energy levels of V$^{3+}$ ions in AV$_2$O$_4$ as functions of
molecular field $h_z$. $J_z=- 2$ is a ground state for all values of
$h_z$ (solid blue bold line). The transitions are possible only to
excited states with $J_z=- 1$ (solid red thin lines).} \label{fig3}
\end{figure}
 When the tetrahedra are
flattened as  observed in the experiments, i.e. when $c<0$,  for any
strength  of the   molecular field $h_z$ the ground state of $H_1$
has  $J_z=\pm 2$ (the sign of $J_z$ depends on the sign of  $h_z$).
For definiteness, we consider spin-down sublattice ($h_z > 0$). The
resulting energy levels  are presented in the Fig.\ref{fig3}. The
magnetic excitations of the local Hamiltonian $H_1$ are transitions
from the ground state to eight excited states of a single ion. These
excitations can be described by boson operators $p_{\nu}^{\dagger}$
-- each pseudoboson describes the
 transition from the ground state $|0\rangle$ to the excited state
 $|\nu\rangle$.
The single ion Hamiltonian $H_1$   is  diagonal in  terms of
pseudoboson operators:
\begin{equation}
H_1=\sum_i \sum_{\nu =1}^8 \varepsilon_{\nu} p_{\nu i}^{\dagger}p_{\nu i},
\end {equation}
where $\varepsilon_{\nu}$ is the
energy difference between the excited state $|\nu \rangle$ and the
ground state of $H_1$:
\begin{equation}
\varepsilon_{\nu} = E_{\nu}-E_0.
\end{equation}

We next introduce the representation for spin operators in terms of
pseudobosons $p_{\nu}^{\dagger}$. The representation for spin ${{\bf
S}}_p$ on the sublattice $p$ can be written as:
\begin{eqnarray}
\label{szgeneral} \begin{array}{ll}
 S_p^z
=&\langle 0|S_z| 0\rangle+ \sum_{\nu =1}^8 \langle
{\nu}|S_z|0\rangle (p^{\dagger}_{\nu}+p_{\nu})+\\
 &\sum_{\nu =1}^8 (\langle {\nu}|S_z|{\nu}\rangle
-\langle 0|S_z|0\rangle )p_{\nu}^{\dagger}p_{\nu}
\end{array}
\end{eqnarray}
\begin{eqnarray}\begin{array}{ll}
S_p^{\pm} =& \sum_{\nu =1}^8 (\langle {\nu}|S^{\pm}|0 \rangle
p_{\nu}^{\dagger} +\langle 0|S^{\pm}|{\nu}\rangle p_{\nu})
\label{spmgeneral}
\end{array}
\end{eqnarray}

\begin{figure}
\centerline{\epsfig{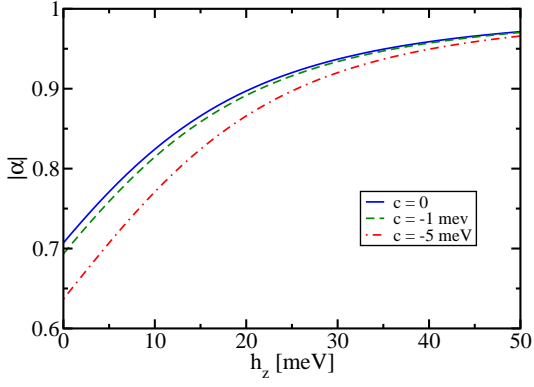}}
 \caption{(Color online) $\alpha$ as a function of molecular field $h_z$ for different
values of the tetragonal field $c$.} \label{fig3a}
\end{figure}

If the ground state of $H_1$ corresponds to $J_z=-2$, it follows
from (\ref{spmgeneral}) that only  states with  $J_z=-1$ contribute
to $S_p^{\pm}$. There are  two such states, $|1\rangle$ and
$|2\rangle$.
 In the basis of $|L'_z,S_z\rangle$ they
have the following structure:
\begin{eqnarray}
\begin{array}{l}
|1\rangle=\sqrt{1-\alpha^2}|-1,0\rangle
+\alpha|0,-1\rangle\\
|2\rangle=\alpha|-1,0\rangle -\sqrt{1-\alpha^2}|0,-1\rangle ~,
\end{array}
\label{states}
\end{eqnarray}
\noindent where we denote $\alpha$ as the weight of the state with
$L'_z= 0$, and $S_z =-1$ in the $|1\rangle$ state.
The  partial weight  of orbital and spin contributions is determined by the
 competition between the spin-orbit coupling, the anisotropy energy
 and the  molecular exchange field.
In Fig.~\ref{fig3a} we present a dependence of $\alpha$ on the
molecular exchange field $h_z$ at different values of the tetragonal
field $c$, keeping the SO coupling constant $\lambda$ fixed. For all
values of tetragonal field parameter $\alpha$ shows similar field
dependence: it increases with the increase of $h_z$. This happens because
the molecular field $h_z$, acting on the spin $S_i$, is effectively
created only by the nearest neighbors on the $xy$-chain, which are
antiferromagnetically aligned to $S_i$. Therefore, when $h_z$
increases, the flipping of the spin costs more energy and the
 transitions with $\Delta S_z= 1$ are suppressed. The low
lying excitations become more of an orbital character,  which can be
seen in the increase of the weight of the transitions with $\Delta
L_z'= 1$.

As only two excited states of an ion are relevant to the spin-wave
analysis, we consider the local excitations described by two
pseudoboson
 operators $p_1^{\dagger}$ and
$p_2^{\dagger}$,  which take an ion from the ground state to states
$|1\rangle$ and $|2\rangle$, respectively.
The explicit form of extended Holstein-Primakoff transformation for
spin ${\bf S}_p$ in terms $p_1$ and $p_2$  is given by:
\begin{eqnarray}
\label{szext} \begin{array}{ll}
 S_p^z
=&\langle 0|S_z|0\rangle+\\&(\langle
1|S_z|1\rangle -\langle 0|S_z|0\rangle )p_1^{\dagger}p_1+\\
&(\langle 2|S_z|2\rangle -\langle 0|S_z|0\rangle )p_2^{\dagger}p_2
\end{array}
\end{eqnarray}

\begin{eqnarray}\begin{array}{ll}
S_p^{\pm} =& <1|S^{\pm}|0>p_1^{\dagger}+<2|S^{\pm}|0>p_2^{\dagger}
+\\ &<0|S^{\pm}|1>p_1+<0|S^{\pm}|2>p_2 \label{spmext}
\end{array}
\end{eqnarray}
\noindent  It is also useful to rewrite these expressions,
(\ref{szext}) and (\ref{spmext}),  using the definition of $\alpha$ (for
spin-down):
\begin{eqnarray}
\label{szext1}
 S_p^z
&=&-1+(1-\alpha^2)p^{\dagger}_1p_1+\alpha^2p^{\dagger}_2p_2\\
S_p^{+}&=&\sqrt{(1-\alpha^2)2}p_1^{\dagger}+\alpha\sqrt{2}p_2^{\dagger}\\
S_p^{-}&=&\sqrt{(1-\alpha^2)2}p_1+\alpha\sqrt{2}p_2 \label{spmext1}
\end{eqnarray}

The interactions between localized excitations are described by
$H_2$.  The excitation spectrum is obtained
in a similar way as in the spin wave analysis for the ROO state.
After diagonalization, the total Hamiltonian $H=H_1+H_2$ can be written
in the same form as
Eqs.(\ref{hdiag}), however now the  index $n$ runs from 1 to 16 and the modes
have complex spin--orbital character. The
details of calculation are given in Appendix B.

\begin{figure}
\centerline{\epsfig{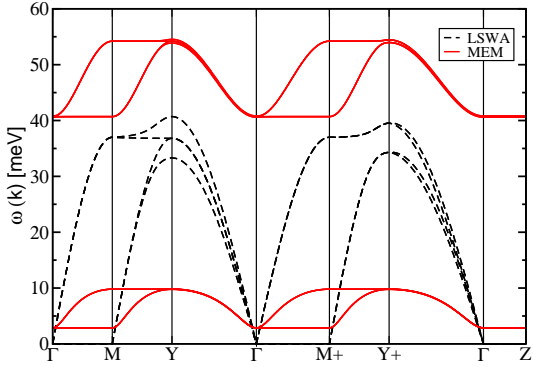}}
 \caption{Magnetic excitation spectrum. Red solid lines correspond
 to magnetic excitations in case of unquenched orbital angular momentum,
 $ L\neq 0$, obtained in the framework of the magnetic exciton
 model (MEM).
We use the following parameters: $J_{xy}=18.5$ meV,
$J'=0.1J_{xy}$, $c=-1$ meV, and  $\lambda=13$ meV. The black
dashed lines correspond to pure spin waves,
 $ L= 0$, obtained in the linear spin wave approximation (LSWA).The
spectra are plotted along the same high symmetry directions as in
Fig.\ref{fig2}.} \label{fig4}
\end{figure}

\begin{figure}
\centerline{\epsfig{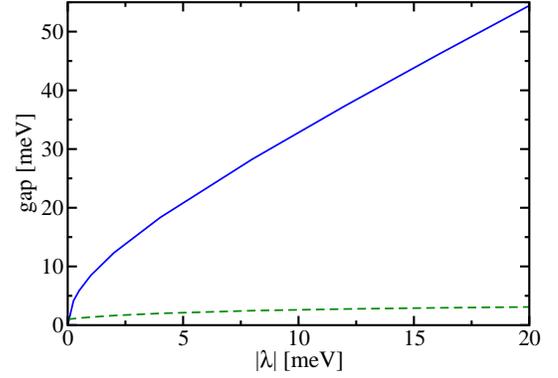}}
 \caption{(Color online) Anisotropy  $\Delta_a$ (dashed line) and optical
 $\Delta_o$ (solid line) gaps
 as a function of spin-orbit coupling constant $\lambda$.} \label{fig4a}
\end{figure}

The numerically calculated magnetic excitation spectrum  for the COO
state is presented in Fig.\ref{fig4}. For comparison, we also
plotted there the spin-wave dispersion for the ROO state.
 We used the
following parameters: $J_{xy}=18.5$ meV, $J'=0.1J_{xy}$, $c=-1$ meV,
and  $\lambda=13$ meV  (third neighbors interactions are not
included here). As we see from Fig.\ref{fig4}, the excitation
spectrum consists  of 8 different branches, each of them is doubly
degenerate. The excitations may be divided into two groups: four low
lying branches with rather small gap $\Delta_a\sim 3$ meV, and four
optical branches, with the gap $\Delta_o\sim 35$ meV. As
$\Delta_a\ll\Delta_o$, the lying branches are quasi-acoustic modes.
As we see in Fig.\ref{fig4}, the lowest mode along the direction
$\Gamma =(0,0,0)\rightarrow M=(\pi/4,\pi/4,0)$ is the dispersionless
mode. This flat mode is a lifted "zero energy mode" of the ROO
state. Here the zero mode is lifted due to the combined effect of
the anisotropy term and the SO coupling. The tetragonal distortion
favors the non-zero value of the $z$-component of orbital angular
momentum, which in turn, selects
 the local spin  quantization axis in such a way, that the $z$
 component of spin is non-zero.
 We found that the gap, $\Delta_a$ is mainly determined by the
anisotropy term and only weakly depends on the SO coupling (see
Fig.\ref{fig4a}).

However, we  note once again, that, in principle, the lifting of
zero energy mode can appear  also in ROO state if single ion
magnetic anisotropy is taken into account. We caution that the size
of these two different anisotropy gaps can be of the same order, and
therefore, it will be rather difficult to distinguish between them.

As  a remark,  we  would like to mention, that such lifted zero
energy modes can be detected by an inelastic neutron scattering, and
have been indeed recently observed in the frustrated Kagome lattice
antiferromagnet KFe$_3$(OH)$_6$(SO$_4$)$_2$~\cite{matan}.

There  is another significant  difference between the magnetic
excitation spectrum for COO and ROO states: as one can see from
Fig.\ref{fig4}, the bandwidth of quasi-acoustic modes in the COO
state is strongly reduced compared to the bandwidth of pure spin
waves in the ROO state.
 This reduction is the effect of the mixing between
 orbital and pure spin  excitations
 in the spectra of the COO state. Pure orbital
excitations are non-dispersive, because they come form local interactions.
 The presence of the
orbital component in the spin-wave spectrum then obviously leads to
the reduction of the  bandwidth.

We believe that this large reduction of the bandwidth can be  seen
experimentally, in the neutron scattering measurements of the
magnetic excitation spectrum. Such  experimental results would
discriminate between two types of orbital ordering, COO and ROO.
Unfortunately, at present these experiments are difficult to perform
 because to measure  the full spin-wave spectrum
  one needs single crystals, which are
 still not available, as far as we know.

Let us now discuss the optical branches. These modes exist only in
the COO state, and they arise from  the hybridization between
orbital angular momentum non-dispersive levels and dispersive spin
branches due to the spin-orbit coupling.
 In Fig.\ref{fig3a} we show how the optical gap, $\Delta_o$,
 depends
on the strength  of the spin-orbit coupling. One can see that at
$\lambda =0$, $\Delta_o$ is zero, but it rapidly increases with
increasing $\lambda$.

 To summarize, the magnetic excitation spectrum
for the COO state has  two separated  branches: quasi acoustic modes
and optical modes. Both manifolds are gapped, but the gaps,
$\Delta_a$ and $\Delta_o$ have   different origins: $\Delta_a$ is
set by the anisotropy term, while $\Delta_o$ is set by the
spin-orbit coupling and is much larger.

These  two gaps  could be, in principle,  determined by inelastic
 neutron or Raman scattering even in powder samples of ZnV$_2$O$_4$.
They are also easily  distinguishable  from the well-known Haldane
gap, $\Delta_H$, which is a characteristic feature of the
antiferromagnetic $S=1$ chains. $\Delta_H$, $\Delta_a$ and
$\Delta_o$ have different temperature dependence: the optical gap,
$\Delta_o$, does not depend on the temperature, as it is determined
by the relativistic spin-orbit interaction; the anisotropy gap,
$\Delta_a$, is non-zero only below the temperature of the structural
transition, $T<T_s$; and
 the Haldane gap should disappear below $T_N$, when a long range
 antiferromagnetic ordering emerges.

At all temperatures, the optical gap, $\Delta_o$, is the largest, so
we compare only the anisotropy and the Haldane gaps.
 At the lowest temperatures, $T<T_N$, the Haldane gap vanishes, and
the lowest mode will have a gap equal to an anisotropy gap
$\Delta_a$.  At intermediate  temperatures, $T_N<T<T_S$,
 the spin excitations are also gapped.  In this temperature range, the
gap is the sum of the anisotropy gap, $\Delta_a$, and the Haldane
gap, $\Delta_H$. The magnitude of the well-developed Haldane gap  is
of the order of $0.4 J_{xy}$, and  is compared to $T_S$, hence it is
very likely that the gap at $T_N<T<T_S$ will be larger then at low
temperatures. This behavior is exotic and, as far as we know,  has
not  been yet observed in  $S=1$ spin-chain systems. At high
temperatures, $T>T_S$, the sharp gap is washed out by thermal
fluctuations, but the spectrum can still divided into acoustic and
optical branches.

\section{Conclusion}

We presented in this paper a detail analysis of the magnetic
excitations of vanadium spinels, whose low-temperature tetragonal
phase  can be modeled to a high accuracy by one-dimensional spin
chains with weak inter-chain interaction.
 The formation of antiferromagnetic spin chains
 on the highly frustrated pyrochlore lattice is by itself non-trivial phenomena.
 This can happen only
because vanadium ions also possess an orbital degree of freedom, and
the orbital modulation of the spin exchange partially lifts the
geometrical degeneracy of the underlying lattice.

We considered two different  ground states: {\it i)} the one with
the real orbital ordering, ROO, and {\it ii)}  the one with the
complex orbital ordering, COO. We  found that the excitation spectra
in these two cases are qualitatively different.  The spectrum for
the COO state consists of low lying quasi-acoustic modes with small
anisotropy gap and  optical branches with zone-center gap determined
by the spin-orbit coupling.  The spectrum for the ROO state  has
only quasi acoustic modes. Within the superesxchange model considered in the
present study the spectrum is gapless, however, in reality we expect an
anisotropy gap also for this state.

The bandwidth of the quasi-acoustic modes in the COO state is strongly
reduced compared to the ones for the ROO state, due to contributions from
orbital $L$-modes. Because the spectra are so  different,
we argue that an effective way to determine experimentally the
symmetry of orbital ordering in vanadium spinels is to measure their
magnetic excitation spectrum.

We gratefully acknowledge discussions with P. Fulde, A. Chubukov, S.
Di Matteo ,  A. Loidl, H. Takagi, R. Valenti and A. Yaresko.

\subsection{Appendix A}

In this appendix we present the
 expressions for the matrices ${\bf A}({\bf
k})$ and ${\bf B}({\bf k})$ in the Hamiltonian (\ref{eakoma}).

The diagonal elements of the matrix ${\bf A}({\bf k})$ are given by
\begin{eqnarray*}
{\bf A}_{pp}({\bf k}) &=& 2J_{xy}+2J_3 \cos2(k_x-k_y), \quad p = 1, 3, 5, 7 \\
{\bf A}_{pp}({\bf k}) &=& 2J_{xy}+2J_3 \cos2(k_x+k_y), \quad p = 2, 4, 6, 8 \\
\end{eqnarray*}

Nonzero matrix elements of ${\bf A}({\bf k})$ are given by
\begin{eqnarray*}
{\bf A}_{14}({\bf k}) &=&J'\,e^{-i(k_x-k_z)}\\
{\bf A}_{16}({\bf k}) &=&J'\,e^{-i(k_y+k_z)}\\
{\bf A}_{23}({\bf k}) &=&J'\,e^{-i(k_x+k_z)}\\
{\bf A}_{25}({\bf k}) &=&J'\,e^{i(k_y+k_z)}\\
{\bf A}_{38}({\bf k}) &=&J'\,e^{i(k_y-k_z)}\\
{\bf A}_{47}({\bf k}) &=&J'\,e^{-i(k_y-k_z)}\\
{\bf A}_{58}({\bf k}) &=&J'\,e^{-i(k_x-k_z)}\\
{\bf A}_{67}({\bf k}) &=&J'\,e^{-i(k_x+k_z)}\\
\\
{\bf A}_{pq}({\bf k}) &=& {\bf \tilde A}_{qp}({\bf k})\\
\label{Ak}
\end{eqnarray*}

Nonzero matrix elements of ${\bf B}({\bf k}) $ are given by

\begin{eqnarray*}
{\bf B}_{13}({\bf k}) &=&2J_{xy} \cos(k_x-k_y) \\
{\bf B}_{24}({\bf k}) &=&2J_{xy} \cos(k_x+k_y) \\
{\bf B}_{57}({\bf k}) &=&2J_{xy} \cos(k_x-k_y) \\
{\bf B}_{68}({\bf k}) &=&2J_{xy} \cos(k_x+k_y) \\
\\
{\bf B}_{12}({\bf k}) &=&J'\,e^{i(k_y+k_z)}\\
{\bf B}_{18}({\bf k}) &=&J'\,e^{i(k_x-k_z)}\\
{\bf B}_{27}({\bf k}) &=&J'\,e^{i(k_x+k_z)}\\
{\bf B}_{34}({\bf k}) &=&J'\,e^{-i(k_y-k_z)}\\
{\bf B}_{36}({\bf k}) &=&J'\,e^{-i(k_x+k_z)}\\
{\bf B}_{45}({\bf k}) &=&J'\,e^{-i(k_x-k_z)}\\
{\bf B}_{56}({\bf k}) &=&J'\,e^{i(k_y+k_z)}\\
{\bf B}_{78}({\bf k}) &=&J'\,e^{-i(k_y-k_z)}\\
\\
{\bf B}_{15}({\bf k}) &=&2J_3 \cos2(k_x-k_z) \\
{\bf B}_{26}({\bf k}) &=&2J_3 \cos2(k_y-k_z) \\
{\bf B}_{37}({\bf k}) &=&2J_3 \cos2(k_x+k_z) \\
{\bf B}_{48}({\bf k}) &=&2J_3 \cos2(k_y-k_z) \\
\\
{\bf B}_{pq}({\bf k}) &=& {\bf \tilde B}_{qp}({\bf k})\\
\label{Bk}
\end{eqnarray*}

\subsection{Appendix B}

We next consider the modifications of the matrices for the exciton
model.

 In the exciton model, the base object is an enlarged set of pseudoboson
 operators  ${\bf \hat a}(\bf k)= ({\bf a_1}(\bf k), {\bf a_2}({\bf k}))$,
 whose components  ${\bf a_1}({\bf k})$  and ${\bf a_2}({\bf k})$ describe
 transitions to  first and  second excited levels. These two
 vectors are analogous  to ${\bf a}({\bf k})$ for pure spin wave model.
As a result, the matrix ${\bf A}({\bf k})$ is enlarged and becomes a
$16 \times 16$ matrix ${\bf \hat A}({\bf k})$.
 Its components are given by
\begin{eqnarray}
{\bf \hat A}({\bf k})= \left(\begin{array}{cc} (1-\alpha^2)\mathcal
A({\bf k}) + \varepsilon_1 &
\alpha\sqrt{1-\alpha^2} \mathcal A({\bf k})  \\
\alpha\sqrt{1-\alpha^2}\mathcal A({\bf k})  & \alpha^2 \mathcal
A({\bf k}) + \varepsilon_2
\end{array} \right),
\label{tenzora}
\end{eqnarray}
where
\begin{equation}
\mathcal A_{pq}({\bf k}) =\left\{ \begin{array}{ll}
 {\bf A}_{pq}({\bf k}) & p \neq q \\
 0                     & p = q
 \end{array} \right.
\end{equation}
Here $p$ and $q$ runs from 1 to 8. The diagonal matrix elements of
$\mathcal A_{pq}({\bf k})$ are  zero (we consider only the case $J_3 =0$),
because the diagonal contribution is already included in $\varepsilon_{\nu}$.

Similarly,  we can obtain the expression for the matrix ${\bf \hat
B}({\bf k})$:
\begin{eqnarray}
\left(\begin{array}{cc}
(1-\alpha^2){\bf B}({\bf k})  & \alpha\sqrt{1-\alpha^2} {\bf B}({\bf k})  \\
\alpha\sqrt{1-\alpha^2}{\bf B}({\bf k})  & \alpha^2 {\bf B}({\bf k})
\end{array} \right).
\label{tenzorb}
\end{eqnarray}


\begin{thebibliography}{99}
\bibitem{notes1}
We would like to note that the coupling between third neighbors
arise not only due to the direct exchange, but also due to the
various super-exchange paths through oxygen ions. In some
pyrochlore compounds this contribution can be rather significant,
however, here, for simplicity, we did not take this coupling into
account.
\end{thebibliography}
\end{document}